%% file: PA006272.tex
\documentclass[11pt,draft,onecolumn]{IEEEtran}
\usepackage{amsmath,amssymb,graphicx,latexsym,citesort,extraipa,mathrsfs}

\newtheorem{theorem}{Theorem}
\newtheorem{theorema}{Theorem}

\newtheorem{theoremb}{Theorem}

\newtheorem{corollary}{Corollary}[theorem]
\newtheorem{lemma}{Lemma}
\newtheorem{proposition}{Proposition}
\newtheorem{definition}{Definition}

\newcommand{\mbf}[1]{\mathbf{#1}}

\newcommand{\set}[1]{\mathscr{#1}}

\newcommand{\SF}[1]{\set{F}(\set{#1})}
\newcommand{\shift}[1]{T_{\set{#1}}}
\newcommand{\shiftn}[2]{T_{\set{#1}}^{#2}}
\newcommand{\shiftns}[3]{T_{\set{#1}^{#3}}^{#2}}
\newcommand{\dynsys}[3]{\ensuremath{(\set{#1}}, \ensuremath{\SF{#1}}, \ensuremath{#2}, \ensuremath{\shift{\set{#3}})}}

\graphicspath{{.}{./figs/}}

\begin{document}

\title{Word-Valued Sources: an Ergodic Theorem, an AEP and the Conservation of Entropy}
%\markboth{Draft: 18-08-2008}{Timo \MakeLowercase{\em et al.}: An Ergodic Theorem for Word-Valued Sources}
\author{R. Timo~\IEEEmembership{Member,~IEEE}, K. Blackmore~\IEEEmembership{Member,~IEEE}, \\ and L. Hanlen~\IEEEmembership{Member,~IEEE}
\thanks{R. Timo is with the Institute for Telecommunications Research at the University of South Australia (e-mail: \texttt{roy.timo@ unisa.edu.au}). K. Blackmore is with the Australian National University (e-mail: \texttt{kim.blackmore@anu.edu.au}). L. Hanlen is with NICTA and the Australian National University (e-mail: \texttt{leif.hanlen@nicta.com.au}). This work was funded by NICTA and the Australian Research Council under the Discovery Grant DP0880223. NICTA is funded by the Australian Government's Backing Australia's Ability initiative, in part through the Australian Research Council. Some of the material in this paper was presented at the $2007$ IEEE International Conference on Networks, Adelaide, Australia, and the $2007$ Australasian Telecommunication Networks and Applications Conference, Christchurch, New Zealand.}}
%\pubid{0000--0000/00\$00.00~\copyright~200X IEEE}

\maketitle
\input{abstract}

\newpage
\input{section1}

\input{section2}

\input{section3}
\input{section4}

\input{section5}

\input{section6}

\input{section7}

\input{acknowledgements}

%\appendix
%\input{appendix}
\bibliographystyle{IEEEtran}
%\bibliography{Library}

% Generated by IEEEtran.bst, version: 1.13 (2008/09/30)

\end{document}

%% file: abstract.tex
\begin{abstract}
A word-valued source $\mathbf{Y} = Y_1,Y_2,\ldots$ is discrete random process that is formed by sequentially encoding the symbols of a random process $\mathbf{X} = X_1,X_2,\ldots$ with codewords from a codebook $\mathscr{C}$. These processes appear frequently in information theory (in particular, in the analysis of source-coding algorithms), so it is of interest to give conditions on $\mathbf{X}$ and $\mathscr{C}$ for which $\mathbf{Y}$ will satisfy an ergodic theorem and possess an Asymptotic Equipartition Property (AEP). In this correspondence, we prove the following: (1) if $\mathbf{X}$ is asymptotically mean stationary, then $\mathbf{Y}$ will satisfy a pointwise ergodic theorem and possess an AEP; and, (2) if the codebook $\mathscr{C}$ is prefix-free, then the entropy rate of $\mathbf{Y}$ is equal to the entropy rate of $\mathbf{X}$ normalized by the average codeword length.
\end{abstract}

%% file: section1.tex
% Section 1: Introduction
\section{Introduction}\label{Sec:Sec1}

The following notion of a word-valued source appears frequently in source-coding theory~\cite{Nishiara-May-2000-A,Goto-Oct-2003-A,Ishida-Dec-2003-A,Ishida-Dec-2006-A}. Suppose that $\set{A}$ and $\set{B}$ are discrete-finite alphabets and $\mbf{X} = X_1,X_2,\ldots$ is an $\set{A}$-valued random process. Let $\set{C}$ be a codebook whose codewords take symbols from $\set{B}$ and have different lengths, and let $f: \set{A} \rightarrow \set{C}$ be a mapping. The word-valued source generated by $\mbf{X}$ and $f$ is the $\set{B}$-valued random process $\mbf{Y} = f(X_1),f(X_2),\ldots$, which is formed by sequentially encoding the symbols of $\mbf{X}$ with $f$ and concatenating (placing end-to-end) the resulting codewords.

It is of fundamental interest to give broad conditions on $\mbf{X}$, $f$ and $\set{C}$ for which $\mbf{Y}$ is guaranteed to possess an Asymptotic Equipartition Property (AEP). A common approach to this type of problem is to determine when the random processes of interest are stationary, after which the classic Shannon-McMillan-Breiman Theorem~\cite[Thm. 15.7.1]{Cover-1991-B} may be used to achieve an AEP. However, this approach is not particularly useful for word-valued sources: for most choices of $f$ and $\set{C}$, $\mbf{Y}$ will not be stationary -- even when $\mbf{X}$ is stationary. Thus, the primary focuss of this paper is to give broad conditions for an AEP without direct recourse to stationarity and the Shannon-McMillan-Breiman Theorem.

Nishiara and Morita~\cite[Thms. 1 $\&$ 2]{Nishiara-May-2000-A} derived an AEP as well as a conservation of entropy law for $\mbf{Y}$ when $\mbf{X}$ is independent and identically distributed (i.i.d.), $f$ is a bijection and $\set{C}$ is prefix-free. (A codebook is said to be prefix-free if no codeword is a prefix of another codeword~\cite[Chap. 5]{Cover-1991-B}.) These results were later extended from the i.i.d. case to the more general stationary and ergodic case by Goto {\em et al.} in~\cite[Thm. 2]{Goto-Oct-2003-A}. We further generalize the results of~\cite{Nishiara-May-2000-A,Goto-Oct-2003-A} to the setting where $\mbf{X}$ is Asymptotically Mean Stationary (AMS), $f$ is a bijection and $\set{C}$ is prefix-free. (This AMS condition is a weaker version of the stationary condition that permits short-term non-stationary properties~\cite{Gray-1987-B}.)
As we will see, the resulting AEP and entropy-conservation law do not retain the simplicity of those results reported in~\cite{Nishiara-May-2000-A,Goto-Oct-2003-A} for stationary and ergodic $\mbf{X}$; namely, both extensions are ineluctably linked to an ergodic-decomposition theorem.

In contrast to the aforementioned results for prefix-free codebooks, very little is know about word-valued sources generated by codebooks without the prefix-free property. In~\cite{Nishiara-May-2000-A}, Nishiara and Morita derived an upper bound for the sample-entropy rate of $\mbf{Y}$ when $\mbf{X}$ is an i.i.d. process and $\set{C}$ is not prefix-free. This upper bound was later supplemented with a non-matching lower bound by Ishida {\em et al.} in~\cite{Ishida-Dec-2006-A}. These bounds, however, fell short of proving an AEP. We prove an ergodic theorem as well as an AEP for $\mbf{Y}$ when $\mbf{X}$ is AMS and $\set{C}$ is arbitrary; and, in doing so, we resolve the open problem reported in~\cite{Nishiara-May-2000-A,Goto-Oct-2003-A,Ishida-Dec-2006-A}.

Our results will follow from a new lemma (Lemma~\ref{Sec:3:Lem:Variable-Length-Shift-AMS}) for AMS random processes. This lemma is an extension of a result by Gray and Saadat~\cite[Cor. 2.1]{Gray-Jan-1984-A}, and it demonstrates that the AMS property is invariant to variable-length time shifts: an AMS random process will remain AMS when it is viewed under different time scales. This invariance property will, in turn, allow us to show that $\mbf{Y}$ is AMS whenever $\mbf{X}$ is AMS -- no matter which $f$ and $\set{C}$ is used. Finally, Gray and Kieffer's AEP for AMS processes~\cite[Cor. 4]{Gray-Oct-1980-A} will provide the desired AEP for $\mbf{Y}$.

An outline of the paper is as follows. We introduce some notation and definitions in Section~\ref{Sec:2}. We present an ergodic theorem (Theorem~\ref{Sec:3:The:Ergodic}) in Section~\ref{Sec:3}, and in Section~\ref{Sec:4} we restate this ergodic theorem using the language of AMS random processes (Theorem~\ref{Sec:4:The:AMS-Word-Valued-Source}). We present an AEP (Theorem~\ref{Sec:5:The:WVS-AEP}) in Section~\ref{Sec:5}. Finally,  Theorems~\ref{Sec:4:The:AMS-Word-Valued-Source} and~\ref{Sec:5:The:WVS-AEP} are proved in  Sections~\ref{Sec:6} and~\ref{Sec:7} respectively.

%% file: section2.tex
\section{Dynamical Systems $\&$ Word-Valued Sources}\label{Sec:2}

The notion of ``time'' is problematic for the development of word-valued sources. In particular, each symbol $X_i$, $i = 1,2,\ldots$, will produce multiple symbols (a codeword) $f(X_i)$; thus, $\mbf{X}$ and $\mbf{Y}$ are naturally defined by different time scales. We simplify notation for these different time scales by using various shift transformations to model the passage of time. A brief review of these transformations and the resulting dynamical systems is given in this section -- a complete treatment can be found in~\cite{Gray-1987-B} and~\cite{Shields-1996-B}. After this review, we formally define word-valued sources.

\input{./section2a}

\input{./section2b}

%% file: section2a.tex
\subsection{A Dynamical Systems Model for $\mbf{X}$}

Let us first introduce some notation. Suppose that $\set{A}$ is a discrete-finite alphabet. For any natural number $n$ (i.e. $n \in \{1,2,\ldots\}$), let
\begin{equation*}
\set{A}^n = \underbrace{\set{A} \times \set{A} \times \cdots \times \set{A}}_{n}
\end{equation*}
denote the $n$-fold Cartesian product of $\set{A}$, and let\footnote{When $n=1$, we shall omit the superscript for brevity, e.g., $a^1 = a$ and $\set{A}^1 = \set{A}$.}  $a^n = a_1,a_2,\ldots,a_n$ denote an arbitrary $n$-tuple from $\set{A}^n$. (These notation conventions will apply to the Cartesian product of every discrete-finite alphabet used in this paper.)

Now suppose that $\mbf{X} = X_1,X_2,\ldots$ is an $\set{A}$-valued random process that is characterised by a sequence of joint probability distributions
\begin{equation}\label{Sec:2:Eqn:Distribution}
p^{(n)}(a^n) = \text{Pr}\big(X_1 = a_1,\ X_2 = a_2,\ldots,\ X_n = a_n\big)\ ,\ \ n = 1,2,\ldots\ ,
\end{equation}
for which the consistency condition
\begin{equation}\label{Sec:2:Eqn:Consistency}
p^{(n)}(a_1,a_2,\ldots,a_n) = \sum_{\tilde{a} \in \set{A}} p^{(n+1)}\big(a_1,a_2,\ldots,a_n,\tilde{a}\big) \ ,\ \ n = 1,2,\ldots\ ,
\end{equation}
is satisfied. Instead of characterising $\mbf{X}$ with the sequence of joint distributions given in~\eqref{Sec:2:Eqn:Distribution}, we may use a dynamical system without loss of generality. A brief review of this fact is as follows.

Let $\set{X} = \set{A} \times \set{A} \times \cdots$ denote the set of all sequences with elements from $\set{A}$, and let $\mbf{x} = x_1,x_2,\ldots$ denote an arbitrary member of $\set{X}$. Now let
\begin{equation*}
\big[a^n\big] = \big\{\mbf{x}\in\set{X} : x_1 = a_1,\ x_2 = a_2,\ldots,x_n = a_n\big\}
\end{equation*}
denote the cylinder set determined by an $n$-tuple $a^n \in \set{A}^n$, and define $\SF{X}$ to be the $\sigma$-field of subsets of $\set{X}$ that is generated by the collection of all cylinder sets. Let $\shift{X} : \set{X} \rightarrow \set{X}$ be the left-shift transform that is defined by
$\shift{X}(\mbf{x}) = x_2,x_3,\ldots$. For integers $n \geq 0$, let\footnote{If $n = 0$, define $\shiftn{X}{0}(\mbf{x}) = \mbf{x}$.}
\begin{align*}
\shiftn{X}{n}(\mbf{x})
&= \underbrace{\shift{X}\Big(\shift{X}\big(\cdots \shift{X}(\mbf{x})\cdots\big)\Big)}_{n}\\
&= x_{n+1},x_{n+2},\ldots
\end{align*}
denote the $n$-fold composition of $\shift{X}$, and let
\begin{equation*}
\shiftn{X}{-n}A = \Big\{\mbf{x} \in \set{X}\ :\ \shiftn{X}{n}(\mbf{x}) \in A\Big\}
\end{equation*}
denote the preimage of an arbitrary set $A \in \SF{X}$ under $\shiftn{X}{n}$. Finally, consider the partition
$\mathcal{Q} = \{[a]:\ a \in \set{A}\}$
of $\set{X}$, and define the function $X_{\mathcal{Q}} : \set{X} \rightarrow \set{A}$ by setting $X_{\mathcal{Q}}(\mbf{x}) = a$ if $\mbf{x} \in [a]$. I.e. $X_{\mathcal{Q}}(\mbf{x})$ returns the value of the first symbol, $x_1$, from $\mbf{x}$.

\medskip

\begin{proposition}[\cite{Gray-1987-B,Shields-1996-B}]\label{Sec:2:Prop:1}
{\em If $\mbf{X}$ is an $\set{A}$-valued random process that is characterised by a distribution~\eqref{Sec:2:Eqn:Distribution} for which the consistency condition~\eqref{Sec:2:Eqn:Consistency} holds, then there exists a unique probability measure $\mu$ on $(\set{X},\SF{X})$ such that
$p^{(n)}(a^n) = \mu([a^n])$ for every tuple $a^n \in \set{A}^n$ and every $n = 1,2,\ldots$. In particular, the distribution of the sequence of $\set{A}$-valued random variables $X_{\mathcal{Q}} \circ \shiftn{X}{n}$, $n = 0,1,\ldots$, defined on $(\set{X},\SF{X},\mu)$ matches that of $\mbf{X}$:
\begin{small}
\begin{equation*}
\mu\Bigg(\Big\{\mbf{x} \in \set{X}\ :\   X_{\mathcal{Q}}(\mbf{x}) = a_1,\ X_{\mathcal{Q}}\big(\shift{X}(\mbf{x})\big) = a_2,\ldots, X_{\mathcal{Q}}\big(\shiftn{X}{n-1}(\mbf{x})\big)=a_n \Big\}\Bigg)
= \mu \left(\bigcap_{i=1}^n \shiftn{X}{-i+1}[a_i]\right) = \mu([a^n])\ .
\end{equation*}
The probability measure $\mu$ is called the Kolmogorov measure of the process $\mbf{X}$.
\end{small}
}
\end{proposition}

\medskip

Proposition~\ref{Sec:2:Prop:1} shows that the quadruple \dynsys{X}{\mu}{X} may be used in place of $\mbf{X}$ without loss of generality. We shall use \dynsys{X}{\mu}{X} and $\mbf{X}$ interchangeably.

%% file: section2b.tex
\subsection{A Dynamical System Model for $\mbf{Y}$}

Suppose that $\set{B}$ is a discrete-finite alphabet, $N$ is a natural number, and
\begin{equation*}
\set{B}^* = \bigcup_{i=1}^N \set{B}^i
\end{equation*}
is the set of all $\set{B}$-valued tuples $b^i = b_1,b_2,\ldots,b_i$ whose length $i$ is greater than or equal to $1$ and no more than $N$.  Let $f:\set{A} \rightarrow \set{B}^*$ be a mapping and $\set{C} = \text{\textit{Range}}(f)$. Finally, let $c$ denote an arbitrary member of $\set{C}$ and $|c|$ its length. We call $f$ a \textit{word function}, $\set{C}$ a \textit{codebook}\footnote{By construction, we have that the length $|c|$ of each codeword $c \in \set{C}$ is bound by $1 \leq |c| \leq N$. In practice, however, the restriction to codewords with finite length may not be suitable for all applications~\cite{Nishiara-May-2000-A}.}, and $c$ a \textit{codeword}.

\medskip

\begin{definition}[Word-Valued Source]
Suppose that $\mbf{X}$ is an $\set{A}$-valued random process and $f$ is a word function. The word-valued source $\mbf{Y}$ generated by $\mbf{X}$ and $f$ is defined to be the $\set{B}$-valued random process that is formed by:
\begin{enumerate}
\item[(i)] sequentially coding the symbols $X_i$, $i = 1,2,\ldots,$ with $f$, and
\item[(ii)] concatenating the resulting sequence of codewords: $\mbf{Y} = f(X_1),$ $f(X_2)$, $f(X_3)$, $\ldots$.
\end{enumerate}
\end{definition}

\medskip

For arbitrary $f$, the particular realisation of $\mbf{X}$ may not be uniquely determined by observing $\mbf{Y}$. The following definition describes a class of word functions where $\mbf{X}$ can be uniquely recovered from $\mbf{Y}$.

\medskip

\begin{definition}[Prefix-Free Word Function]
A word function $f$ is said to be prefix free if:
\begin{enumerate}
\item[(i)] $f:\set{A} \rightarrow \set{C}$ is a bijection, and
\item[(ii)] there does not exist two codewords $c$ and $c'$ in $\set{C}$ such that $c_i = c'_i$ for $i = 1,2,\ldots,\min\{|c|,|c'|\}$.
\end{enumerate}
\end{definition}

\medskip

The distribution of the word-valued source $\mbf{Y}$,
\begin{equation*}
q^{(n)}\big(b^n\big) = \text{Pr}\big(Y_1=b_1, Y_2 = b_2, \ldots, Y_n=b_n\big)\ ,\  n = 1,2,\ldots\ ,
\end{equation*}
may be calculated by combining the distribution of $\mbf{X}$ with $f$. With a slight abuse of notation, let $f^{-1}b^n$ denote the set of $n$-tuples $a^n$ where the first $n$ symbols of the $n$ concatenated codewords $f(a_1)$, $f(a_2)$, $\ldots$, $f(a_n)$ are equal to $b^n$; that is,
\begin{equation*}
f^{-1}b^n = \Big\{ a^n \in \set{A}^n\ :\ \phi_n\big(f(a_1),f(a_2),\ldots,f(a_n)\big) = b^n\Big\}\ ,
\end{equation*}
where $\phi_n: \cup_{n \leq m \leq nN} \set{B}^m \rightarrow \set{B}^n$ is the projection defined by $\phi_n(b_1,$ $b_2,$ $\ldots,$ $b_n,$ $b_{n+1},$ $\ldots,$ $b_m)$ $=$ $b_1,$ $b_2,$ $\ldots,$ $b_n$. Using this notation, we have that
\begin{equation}\label{Sec:2:Eqn:Dist-Y-2}
q^{(n)}\big(b^n\big) =
\left\{
  \begin{array}{ll}
    \sum_{a^n \in f^{-1}b^n} p^{(n)}\big(a^n\big), & \hbox{if $f^{-1}b^n \neq \emptyset$ and} \\
    0, & \hbox{otherwise,}
  \end{array}
\right.
\end{equation}
where $\emptyset$ denotes the empty set.

Describing $\mbf{Y}$ directly with~\eqref{Sec:2:Eqn:Dist-Y-2} is rather cumbersome, and it is more convenient to use a dynamical system that is formed by coding \dynsys{X}{\mu}{X} with a sequence-to-sequence coder. To this end, let $\set{Y} = \set{B} \times \set{B} \times \cdots$ denote the collection of all sequences with elements from $\set{B}$, let $\mbf{b} = b_1,b_2,\ldots$ denote an arbitrary member of $\set{Y}$, and let $\SF{Y}$ be the $\sigma$-field of subsets of $\set{Y}$ generated by cylinder sets. Now consider the sequence-to-sequence coder (measurable mapping) $F: \set{X} \rightarrow \set{Y}$ that is formed by setting $F(\mbf{x})$ $=$ $f(x_1)$, $f(x_2)$, $\ldots$.
When $F$ acts on the abstract probability space $(\set{X}$, $\SF{X}$, $\mu)$, it induces a probability measure $\eta$ on $(\set{Y},\SF{Y})$~\cite[Ex. 9.4.3]{Gray-1990-B} \cite[Pg. 80]{Shields-1996-B}. In particular, $\eta$ and $\mu$ are related by
\begin{equation}\label{Sec:2:Eqn:Input-Output}
\eta(A) = \mu\big(F^{-1} A \big)\ ,\ \ A \in \SF{Y}\ ,
\end{equation}
where $F^{-1}A = \{\mbf{x} \in \set{X}$ : $F(\mbf{x}) \in A\}$ denotes the preimage of a set $A \in \SF{Y}$ under $F$. Finally, when $(\set{Y},\SF{Y},\eta)$ is combined with the left-shift transform $\shift{Y}(\mbf{y}) = y_2,y_3\ldots$ and the partition $\{[b]:b \in \set{B}\}$ of $\set{Y}$, the result is a dynamical system model \dynsys{Y}{\eta}{Y} for $\mbf{Y}$. In particular, for each $n = 1,2,\ldots$ and $b^n \in \set{B}^n$, we have that
$\eta\big([b^n]\big)$ $=$ $\mu\big(F^{-1}[b^n]\big)$ $=$ $q^{(n)}\big(b^n\big)$.

Throughout the remainder of this paper, we shall use the following notation: \dynsys{X}{\mu}{X} and $\mbf{X}$ will denote an arbitrary $\set{A}$-valued random process; $f:\set{A} \rightarrow \set{C}$ will denote a word function; $F: \set{X} \rightarrow \set{Y}$ will denote the sequence-to-sequence coder generated by $f$; and, \dynsys{Y}{\eta}{Y} and $\mbf{Y}$ will denote the word-valued source generated by coding \dynsys{X}{\mu}{X} with $F$, where $\mu$ and $\eta$ are related via~\eqref{Sec:2:Eqn:Input-Output}. In addition, we will use $(\set{W}$, $\SF{W}$, $\rho$, $T)$ to represent an arbitrary dynamical system. Here it should always be understood that $\set{W}$ is the sequence space corresponding to some discrete-finite alphabet (an element of which will be written $\mbf{w} = w_1,w_2,\ldots$); $\SF{W}$ is the $\sigma$-field generated by cylinder sets; $\rho$ is a probability measure on $(\set{W},\SF{W})$; and, $T: \set{W} \rightarrow \set{W}$ is an arbitrary measurable mapping. When we are explicitly interested in the special case where $T$ is the left-shift transform, we shall use the notation $\shift{W}(\mbf{w}) = w_2,w_3,\ldots$.

%% file: section3.tex
\section{A Pointwise Ergodic Theorem}\label{Sec:3}

\medskip
\setcounter{theorem}{1}
\begin{theorema}\label{Sec:3:The:Ergodic}
{\em \
\begin{enumerate}
\item[(i)] If the limit
\begin{equation}\label{Sec:2:Eqn:1}
\langle g \rangle(\mbf{x}) = \lim_{n\rightarrow\infty}\frac{1}{n}\sum_{i=0}^{n-1} g \big(\shiftn{X}{i}(\mbf{x})\big)
\end{equation}
exists almost surely with respect to $\mu$ (a.s. [$\mu$]) for every bounded-measurable $g: \set{X} \rightarrow (-\infty,\infty)$,  then the limit
\begin{equation}\label{Sec:2:Eqn:2}
\langle \tilde{g} \rangle(\mbf{y}) = \lim_{m\rightarrow\infty}\frac{1}{m}\sum_{j=0}^{m-1} \tilde{g} \big(\shiftn{Y}{j}(\mbf{y})\big)
\end{equation}
exists a.s. [$\eta$] for every bounded-measurable $\tilde{g}: \set{Y} \rightarrow (-\infty,\infty)$. If $f$ is prefix-free, then the reverse implication also holds.

\item[(ii)] If the limit~\eqref{Sec:2:Eqn:1} exists and takes a constant value a.s. [$\mu$] for every bounded-measurable $g: \set{X} \rightarrow (-\infty,\infty)$, then the limit~\eqref{Sec:2:Eqn:2} exists and takes a constant value a.s. [$\eta$] for every bounded-measurable $\tilde{g}: \set{Y} \rightarrow (-\infty,\infty)$.
\end{enumerate}
}
\end{theorema}

%% file: section4.tex
\section{Asymptotically Mean Stationary Random Processes}\label{Sec:4}

Theorem~\ref{Sec:3:The:Ergodic} may be restated in a more compact form using the language of asymptotically mean stationary random processes. For this purpose, let us recall the following definitions from Gray~\cite{Gray-1987-B}.

Consider a dynamical system \dynsys{W}{\rho}{}, where $T: \set{W} \rightarrow \set{W}$ is an arbitrary measurable mapping. The system is said to be \textit{stationary} if $\rho(A) = \rho(T^{-1}A)$ for every $A \in \SF{W}$. A set $A \in \SF{W}$ is said to be $T$-invariant if $A = T^{-1}A$. The system is said to be \textit{ergodic} if $\rho(A) = 0$ or $1$ for every $T$-invariant set $A$. Finally, the system is said to be \textit{Asymptotically Mean Stationary} (AMS) if the limit
\begin{equation*}%\label{Sec:2:Eqn:3}
\lim_{n\rightarrow\infty}\frac{1}{n}\sum_{i=0}^{n-1} \rho \big(\shiftn{}{-i} A \big)
\end{equation*}
exists for every $A \in \SF{W}$, in which case the set function
\begin{equation*}
\overline{\rho}(A) = \lim_{n\rightarrow\infty}\frac{1}{n}\sum_{i=0}^{n-1} \rho \big(\shiftn{}{-i} A \big)\ ,\ A \in \SF{W},
\end{equation*}
is a stationary probability measure on $(\set{W},\SF{W})$; that is, the system $(\set{W}$, $\SF{W}$, $\overline{\rho}$, $T)$ is stationary. The measure $\overline{\rho}$ is called the \textit{stationary mean} of $\rho$.

For brevity, we will say that the measure $\rho$ is $T$-stationary / $T$-ergodic / $T$-AMS if the corresponding dynamical systems is stationary / ergodic / AMS respectively. The next lemma gives necessary and sufficient conditions for a system to be ergodic and AMS.

\medskip

\begin{lemma}\label{Sec:4:Lem:Equiv-AMS}
{\em \
\begin{enumerate}
\item[(i)] The system \dynsys{W}{\rho}{} is AMS if and only if the limit
\begin{equation}\label{Sec:4:Eqn:AMS-Ergodic-1}
\langle g \rangle(\mbf{w}) = \lim_{n\rightarrow\infty}\frac{1}{n} \sum_{i=0}^{n-1} g\big(T^i(\mbf{w})\big)
\end{equation}
exists a.s. [$\rho$] for every bounded-measurable $g:\set{W}\rightarrow(-\infty,\infty)$.
\item[(ii)] The system \dynsys{W}{\rho}{} is ergodic if and only if the limit~\eqref{Sec:4:Eqn:AMS-Ergodic-1} takes a constant finite value a.s. [$\rho$] for every bounded-measurable $g:\set{W}\rightarrow(-\infty,\infty)$.
\end{enumerate}
}
\end{lemma}

\medskip

The AMS component of Lemma~\ref{Sec:4:Lem:Equiv-AMS} was proved by Gray and Kieffer~\cite[Thm. 1]{Gray-Oct-1980-A}, and the ergodic component follows from the definition of ergodicity~\cite[Sec. 6.7]{Gray-1987-B}. Using Lemma~\ref{Sec:4:Lem:Equiv-AMS}, we may restate Theorem~\ref{Sec:3:The:Ergodic} as follows. A proof of this result can be found in Section~\ref{Sec:6}.

\medskip

\begin{theoremb}\label{Sec:4:The:AMS-Word-Valued-Source}
{\em \ 
\begin{enumerate}
\item[(i)] If $\mu$ is $\shift{X}$-AMS, then $\eta$ is $\shift{Y}$-AMS. 
\item[(ii)] If $f$ is prefix-free, then $\eta$ is $\shift{Y}$-AMS if and only if $\mu$ is $\shift{X}$-AMS.
\item[(iii)] If $\mu$ is $\shift{X}$-ergodic, then $\eta$ is $\shift{Y}$-ergodic.
\end{enumerate}
}
\end{theoremb}

%% file: section5.tex
\section{An Asymptotic Equipartition Property}\label{Sec:5}

In this section, we extend the AEP of~\cite{Nishiara-May-2000-A,Goto-Oct-2003-A,Ishida-Dec-2006-A} to the setting where $\mu$ is $\shift{X}$-AMS and $f$ is arbitrary. Two fundamental features of this extension will be the ergodic-decomposition theorem and the AEP for AMS random processes. We briefly review each of these ideas in Subsections~\ref{Sec:5a} and~\ref{Sec:5b} before stating our main results in Subsection~\ref{Sec:5c}.

\input{./section5a}
\input{./section5b}

\input{./section5c}

%% file: section5a.tex
\subsection{The Ergodic Decomposition Theorem}\label{Sec:5a}

Suppose that $\mbf{W} = W_1,W_2,\ldots$ is a discrete-finite alphabet random process and $\dynsys{W}{\rho}{W}$ is the corresponding dynamical system in the sense of Proposition~\ref{Sec:2:Prop:1}, where $\shift{W}(\mbf{w}) = w_2,w_3,\ldots$ is the left-shift transformation. For each set $A \in \SF{W}$, let $\mathbf{1}_A$ denote its indicator function:
\begin{equation*}
\mathbf{1}_A(\mbf{w}) = \left\{
               \begin{array}{ll}
                 1, & \hbox{ if } \mbf{w} \in A \\
                 0, & \hbox{ otherwise.}
               \end{array}
             \right.
\end{equation*}
When the limit exists, let
\begin{equation*}
\langle \mathbf{1}_A \rangle (\mbf{w}) = \lim_{n\rightarrow\infty} \frac{1}{n} \sum_{i=0}^{n-1} \mathbf{1}_A\Big(\shiftn{W}{i}(\mbf{w}) \Big)
\end{equation*}
denote the relative frequency of the set $A$ in the sequence $\mbf{w}$. Finally, for each bounded-measurable function $g: \set{W} \rightarrow (-\infty,\infty)$, let $\mathbb{E}[\rho,g]$ denote its expected value:
\begin{equation*}
\mathbb{E}\big[\rho,g\big] = \int g(\mbf{w})\ d\rho(\mbf{w})\ .
\end{equation*}

The pair $(\set{W},\SF{W})$ belongs to a family of measurable spaces called standard spaces~\cite[Chap. 2]{Gray-1987-B}. A distinctive property of these spaces is that they possess a countable generating field~\cite[Cor. 2.2.1]{Gray-1987-B}. Let $\set{S}$ be a countable generating field for $(\set{W},\SF{W})$. Now let $G(\set{S})$ denote the collection of sequences $\mbf{w}$ from $\set{W}$ such that the limit $\langle \mathbf{1}_A \rangle(\mbf{w})$ exists for every generating set $A \in \set{S}$. It can be shown that, for each $\mbf{w} \in G(\set{S})$, the set function $P_{\mbf{w}}$ obtained by setting
$P_{\mbf{w}}(A) = \langle \mathbf{1}_A \rangle(\mbf{w})$ induces a unique $\shift{W}$-stationary probability measure $p_{\mbf{w}}$ on $(\set{W},\SF{W})$. Let $E$ denote the set of sequences $\mbf{w}$ from $G(\set{S})$ where the induced $\shift{W}$-stationary probability measure $p_{\mbf{w}}$ is also $\shift{W}$-ergodic:
\begin{equation*}
E = \big\{ \mbf{w} \in \set{W}\ : \ \mbf{w} \in G(\set{S}) \text{ and } p_{\mbf{w}} \text{ is $\shift{W}$-ergodic}\big\}\ .
\end{equation*}
The set $E$ is called the set of \textit{ergodic sequences}. Finally, let $p^*$ be an arbitrary $\shift{W}$-stationary and $\shift{W}$-ergodic probability measure on $(\set{W},\SF{W})$, and for each sequence $\mbf{w} \in \set{W}$ define
\begin{equation*}
\overline{\rho}_{\mbf{w}} = \left\{
                            \begin{array}{ll}
                              p_{\mbf{w}}, & \hbox{if } \mbf{w} \in E \\
                              p^*, & \hbox{otherwise.}
                            \end{array}
                          \right.
\end{equation*}
The collection of probability measures $\{\overline{\rho}_{\mbf{w}}:\ \mbf{w} \in \set{W}\}$ is called the \textit{ergodic decomposition} of $(\set{W},\SF{W})$.

\medskip

\begin{lemma}[AMS Ergodic Decomposition Theorem~\cite{Gray-1987-B,Shields-1996-B}]\label{Sec:5:Lem:Ergodic-Decomposition}
{\em Let $\{\overline{\rho}_{\mbf{w}}:$ $\mbf{w} \in \set{W}\}$ be the ergodic decomposition of $(\set{W},\SF{W})$ and $E$ the set of ergodic sequences. Then,
\begin{enumerate}
\item[(i)] the set $E$ is $\shift{W}$-invariant: $E = \shift{W}^{-1}E$,
\item[(ii)] $\overline{\rho}_{\mbf{w}}(A) = \overline{\rho}_{\shift{W}(\mbf{w})}(A)$ for every set $A \in \SF{W}$ and every sequence $\mbf{w} \in \set{W}$,
\item[(iii)] for any pair $\mbf{w}$ and $\mbf{w}'$, the probability measures $\overline{\rho}_{\mbf{w}}$ and $\overline{\rho}_{\mbf{w}'}$ are either identical or mutually singular.
\end{enumerate}
Additionally, if $\rho$ is $T$-AMS with stationary mean $\overline{\rho}$, then
\begin{enumerate}
\item[(iv)] $\rho(E) = \overline{\rho}(E) = 1$,
\item[(v)] for each set $A \in \SF{W}$
\begin{equation*}
\overline{\rho}(A) = \int \overline{\rho}_{\mbf{w}}(A)\ d \rho(\mbf{w})\ ,
\end{equation*}
\item[(vi)] the limit
\begin{align*}
\langle g \rangle(\mbf{w}) &= \lim_{n\rightarrow\infty} \frac{1}{n}\sum_{i=0}^{n-1} g\big(\shiftn{W}{i}(\mbf{w})\big) = \mathbb{E}\big[\overline{\rho}_{\mbf{w}},g\big]
\end{align*}
holds a.s. [$\rho$] for each bounded-measurable function $g: \set{W} \rightarrow (-\infty,\infty)$.
\end{enumerate}}
\end{lemma} 

%% file: section5b.tex
\subsection{An AEP for AMS Random Processes}\label{Sec:5b}

As before, suppose that $\mbf{W} = W_1,W_2,\ldots$ is a discrete-finite alphabet random process and \dynsys{W}{\rho}{W} is the corresponding dynamical system. For each sequence $\mbf{w} \in \set{W}$, the probability $\rho([w^n])$ is non-increasing in $n$. If $\rho$ is $\shift{W}$-AMS, then Gray and Kieffer's AEP~\cite{Gray-Oct-1980-A} asserts that this decrease is exponential in $n$ on a set of probability one; in particular, the (asymptotic) rate of decent is given by the entropy rate
of the underlying $\shift{W}$-stationary and $\shift{W}$-ergodic probability measure $\overline{\rho}_{\mbf{w}}$ from the ergodic decomposition theorem. A formal statement of this idea is given in the next lemma. However, before this lemma is given, we briefly  review the concepts of joint entropy, entropy rate and sample-entropy rate.

The \textit{joint entropy} $H(W^n)$ of the first $n$-random variables $W^n$ from $\mbf{W}$ is defined as~\cite{Cover-1991-B}
\begin{equation*}
H(W^n) = \sum_{w^n} \text{Pr}\big[W^n=w^n\big] \log \frac{1}{\text{Pr}\big[W^n=w^n\big]}\ .
\end{equation*}
With respect to the Kolmogorov measure $\rho$, we define the joint entropy of the first $n$ random variables to be
\begin{equation*}
H_n(\rho) =  \sum_{w^n} \rho\big([w^n]\big) \log \frac{1}{\rho\big([w^n]\big)}\ .
\end{equation*}
From Proposition~\ref{Sec:2:Prop:1}, these functionals are consistent in that $H(W^n)=H_n(\rho)$.
When the limit exists, the \textit{entropy rate} of $\mbf{W}$ is defined as
$\overline{H}(\mbf{W}) = \lim_{n\rightarrow\infty}(1/n)H(W^n)$~\cite[Chap. 4]{Cover-1991-B}. Similarly, we define the entropy rate of $\mbf{W}$ with respect to $\rho$ to be
$
\overline{H}(\rho) = \lim_{n\rightarrow\infty}(1/n)H_n(\rho)
$
when the limit exists. Finally, we define the \textit{sample-entropy rate} of a sequence $\mbf{w} \in \set{W}$ with respect to $\rho$ as
\begin{equation*}
h(\rho,\mbf{w}) = \lim_{n\rightarrow\infty}\frac{1}{n} \log \frac{1}{\rho\big([w^n]\big)}\ ,
\end{equation*}
when the limit exists.

\medskip

\begin{lemma}[Asymptotic Equipartition Property~\cite{Gray-1990-B}]\label{Sec:5:Lem:AMS-AEP}
{\em Let $\{\overline{\rho}_{\mbf{w}}:$ $\mbf{w} \in \set{W}\}$ be the ergodic decomposition of $(\set{W},\SF{W})$. If $\rho$ is $\shift{W}$-AMS with stationary mean $\overline{\rho}$, then there exists a set $\Omega \in \SF{W}$ with probability $\rho(\Omega) = 1$ such that the sample-entropy rate $h(\rho,\mbf{w})$ of any sequence $\mbf{w} \in \Omega$ exists and is given by
\begin{equation}\label{Sec:5:Eqn:Sample-Entropy-Rate-Def}
h(\rho,\mbf{w}) = \varphi(\mbf{w})\ ,
\end{equation}
where $\varphi$ is the $\shift{W}$-invariant function that is
defined by $\varphi(\mbf{w}) = \overline{H}(\overline{\rho}_{\mbf{w}})$.
Furthermore, the entropy rate of $\rho$ exists and is given by
\begin{equation*}
\overline{H}(\rho) = \overline{H}(\overline{\rho}) =  \mathbb{E}\big[\rho,\varphi\big]\ .
\end{equation*}
Finally, if $\rho$ is $\shift{W}$-ergodic, then $h(\rho,\mbf{w}) = \overline{H}(\rho) = \overline{H}(\overline{\rho})$ for every $\mbf{w} \in \Omega$.
}
\end{lemma}

%% file: section5c.tex
\subsection{An AEP for Word Valued Sources}\label{Sec:5c}

We now return to the problem of establishing an AEP for $\mbf{Y}$. From Theorem~\ref{Sec:4:The:AMS-Word-Valued-Source} and Lemma~\ref{Sec:5:Lem:AMS-AEP}, it is clear that $\mbf{Y}$ satisfies an AEP whenever $\mu$ is $\shift{X}$-AMS. It turns out, however, that not only does the limit $h(\eta,\mbf{y})$ exist almost surely, but its value may also be bound from above by the entropy rate of $\mbf{X}$ normalized by the expected codeword length. We formalize this idea in the following theorem.

\medskip

\begin{theorem}\label{Sec:5:The:WVS-AEP}
{\em Let $\{\overline{\mu}_{\mbf{x}}\ :\ \mbf{x} \in \set{X}\}$ be the ergodic decomposition of $(\set{X},\SF{X})$. If $\mu$ is $\shift{X}$-AMS, then $\eta$ is $\shift{Y}$-AMS and there exists a set $\Omega_x \in \SF{X}$ with probability $\mu(\Omega_x) = 1$ such that, for every sequence $\mbf{x} \in \Omega_x$, the sample-entropy rate $h(\eta,F(\mbf{x}))$ of the word-valued sequence $F(\mbf{x}) = f(x_1),f(x_2),\ldots$ exists and is bound from above by
\begin{equation}\label{Sec:2:Eqn:WVS-AEP-1}
h\big(\eta,F(\mbf{x})\big)
\leq
\frac{\overline{H}({\overline{\mu}_{\mbf{x}}})}
{\mathbb{E}\big[\overline{\mu}_{\mbf{x}},l\big]}\ ,
\end{equation}
where $l\ :\ \set{X} \rightarrow \{1,2,\ldots,N\}$ is given by $l(\mbf{x}) = |f(x_1)|$. % and
%\begin{align*}
%\mathbb{E}\big[\overline{\mu}_{\mbf{x}},l\big]
%&=\lim_{n\rightarrow\infty} \frac{1}{n}\sum_{i=0}^{n-1} l\big(\shiftn{X}{i}(\mbf{x})\big)
%=\lim_{n\rightarrow\infty} \frac{1}{n}\sum_{i=1}^{n} |f(x_i)|\ .
%\end{align*}
In addition, if $f$ is prefix free, then the inequality in \eqref{Sec:2:Eqn:WVS-AEP-1} becomes an equality.}
\end{theorem}

\medskip

A proof of Theorem~\ref{Sec:5:The:WVS-AEP} follows in Section~\ref{Sec:7}. The next corollary demonstrates that if $\mbf{X}$ is AMS, then the entropy in each symbol of $\mbf{X}$ is conserved with respect to each stationary and ergodic sub-source from the ergodic-decomposition theorem. This behaviour is consistent with the entropy-conservation laws of variable-to-fixed length source codes~\cite{Jelinek-Nov-1992-A,Savari-Jul-1999-A}.

\medskip

\begin{corollary}\label{Sec:5:The:WVS-AEP:Cor:1}
{\em If $\mu$ is $\shift{X}$-AMS, then the entropy rate of $\eta$ exists and is bound from above by
\begin{equation}
\label{Sec:2:Eqn:WVS-AEP-2}
\overline{H}(\eta)
\leq
\int \frac{\overline{H}(\overline{\mu}_{\mbf{x}}\big)}
{\mathbb{E}\big[\overline{\mu}_{\mbf{x}},l\big]}\ d\mu(\mbf{x})\ .
\end{equation}
In addition, if $f$ is prefix-free, then the inequality in \eqref{Sec:2:Eqn:WVS-AEP-2} becomes an equality.}
\end{corollary}

\medskip

Finally, the next corollary resolves the open problem reported in~\cite{Nishiara-May-2000-A,Goto-Oct-2003-A,Ishida-Dec-2006-A}: if $\mbf{X}$ is stationary and ergodic, then an AEP holds for $\mbf{Y}$.

\medskip

\begin{corollary}\label{Sec:5:The:WVS-AEP:Cor:2}
{\em If $\mu$ is $\shift{X}$-stationary and $\shift{X}$-ergodic, then $\eta$ is $\shift{Y}$-ergodic and
\begin{equation}
\label{Sec:2:Eqn:WVS-AEP-3}
h\big(\eta,\mbf{y}\big)
\leq
\frac{\overline{H}(\mu)}{\mathbb{E}\big[\mu,l\big]}\ \ a.s.\ [\eta]\ .
\end{equation}
In addition, if $f$ is prefix-free, then the inequality in \eqref{Sec:2:Eqn:WVS-AEP-3} becomes an equality.}
\end{corollary}

%% file: section6.tex
\section{Proof of Theorem~1}\label{Sec:6}

The proof of Theorem~\ref{Sec:4:The:AMS-Word-Valued-Source} (and Theorem~\ref{Sec:3:The:Ergodic}) will use Lemmas~\ref{Sec:6:Lem:Bellow} through~\ref{Sec:3:Lem:Variable-Length-Shift-Erogidc}, which are given respectively in Subsections~\ref{Sec:6:Sub:A} through~\ref{Sec:6:Sub:E}. The forward and reverse implications of Theorem~\ref{Sec:4:The:AMS-Word-Valued-Source} are proved in Subsections~\ref{Sec:6:Sub:F} and~\ref{Sec:6:Sub:G} respectively.

\input{./section6a}

\input{./section6b}

\input{./section6c}
\input{./section6d}
\input{./section6e}
\input{./section6f}

\input{./section6g}

%% file: section6a.tex
\subsection{Subsequences, Weighted Sequences $\&$ Density}\label{Sec:6:Sub:A}

Suppose that $\zeta$ $=$ $\zeta_0$, $\zeta_1$, $\zeta_2$, $\ldots$ is a strictly increasing subsequence in the non-negative integers $\mathbb{Z}^*$ $=$ $\{0$, $1$, $2$, $\ldots\}$. Let $\xi$ $=$ $\xi_0$, $\xi_1$, $\xi_2$, $\ldots$ be the \textit{weight sequence} obtained from $\zeta$ by setting
\begin{equation}\label{Sec:6:Eqn:Weight-1a}
\xi_n = \left\{
        \begin{array}{ll}
          1, & \hbox{ if } n = \zeta_k \text{ for some } k = 0,1,\ldots \\
          0, & \hbox{otherwise.}
        \end{array}
      \right.
\end{equation}
When the limit exists, the density $d_{\zeta}$ of $\zeta$ in $\mathbb{Z}^*$ is defined as
\begin{equation}\label{Sec:6:Eqn:Density}
d_{\zeta} = \lim_{n\rightarrow\infty}\frac{1}{n}\sum_{i=0}^{n-1} \xi_i\ .
\end{equation}

The next lemma follows directly from these definitions, e.g., see~\cite[Prop. 1.7]{Bellow-Mar-1985-A}.

\medskip

\begin{lemma}\label{Sec:6:Lem:Bellow}
{\em Suppose that $\zeta$ is a strictly increasing subsequence in $\mathbb{Z}^*$ with density $d_{\zeta} > 0$ and weight sequence $\xi$. For any sequence $\mbf{r} = r_0,r_1,\ldots$ of real numbers, we have that
\begin{equation*}
d_{\zeta}\ \lim_{k\rightarrow\infty}\frac{1}{k} \sum_{j=0}^{k-1} r_{\zeta_j} = \lim_{n\rightarrow\infty}\frac{1}{n}\sum_{i=0}^{n-1} \xi_i\ r_i\ ;
\end{equation*}
that is, the existence of either limit implies the existence of the other.}
\end{lemma}

%% file: section6b.tex
\subsection{Invariant Sets $\&$ Asymptotic Mean Stationarity}\label{Sec:6:Sub:B}

The next lemma gives some equivalence conditions for AMS dynamical systems.

\medskip

\begin{lemma}[Cor. 6.3.4, \cite{Gray-1987-B}; Thm. 2.2, \cite{Kakihara-Sep-2001-A}]\label{Sec:6:Lem:AMS-Equivalent}
{\em For a dynamical system \dynsys{W}{\rho}{}, the following statements are equivalent:
\begin{enumerate}
\item[(i)] $\rho$ is $T$-AMS.
\item[(ii)] There exists a $T$-stationary probability measure $\tilde{\rho}$ on $(\set{W},\SF{W})$ such that $\tilde{\rho}$ asymptotically dominates $\rho$; that is, $\tilde{\rho}(A) = 0$ implies
$
\lim_{n\rightarrow\infty}\rho\big(T^{-n}A\big) = 0\ .
$
\item[(iii)] The limit $\lim_{n\rightarrow\infty}$ $(1/n)$ $\sum_{i=0}^{n-1}$ $g(\shiftn{}{i}\mbf{w})$ exists a.s. [$\rho$] for every bounded-measurable $g: \set{W} \rightarrow (-\infty,\infty)$. (See also Lemma~\ref{Sec:4:Lem:Equiv-AMS}.)
\item[(iv)] There exists a $T$-stationary probability measure $\tilde{\rho}$ on $(\set{W},\SF{W})$ such that $A = T^{-1} A$ and $\tilde{\rho}(A) = 0$ together imply that $\rho(A) = 0$.
\end{enumerate}
}
\end{lemma}

%% file: section6c.tex
\subsection{Stationary, Ergodic $\&$ AMS Sequence Coders}\label{Sec:6:Sub:C}

In Section~\ref{Sec:2}, we defined the word-valued source \dynsys{Y}{\eta}{Y} using a sequence coder $F: \set{X} \rightarrow \set{Y}$. In the proof of Theorem~\ref{Sec:4:The:AMS-Word-Valued-Source}, it will be necessary to determine when such a sequence coder will transfer stationary / ergodic / AMS properties from the input to the output. For this purpose, we now review the notions of stationary, ergodic and AMS sequence coders.

Suppose that \dynsys{W}{\rho_{\alpha}}{\alpha} and \dynsys{U}{\rho_{\beta}}{\beta} are dynamical systems, where $\set{W}$ and $\set{U}$ are sequence spaces corresponding to some discrete-finite alphabets; $\SF{W}$ and $\SF{U}$ are $\sigma$-fields generated by cylinder sets; $T_{\alpha}: \set{W} \rightarrow \set{W}$ and $T_\beta : \set{U} \rightarrow \set{U}$ are arbitrary measurable maps; $G : \set{W} \rightarrow \set{U}$ is a sequence coder; $\rho_{\alpha}$ is a probability measure on $(\set{W},\SF{W})$; and, $\rho_{\beta}$ is induced by $G$
\begin{equation*}
\rho_{\beta}\big(A\big) = \rho_{\alpha}\big(G^{-1}A\big)\ ,\ A \in \SF{U}\ .
\end{equation*}
%The systems are said to be \textit{connected} by $G$.

The sequence coder $G$ also induces a probability measure $\rho_{\alpha\beta}$ on the product space\footnote{We use $\SF{W} \times \SF{U}$ to denote the product $\sigma$-field induced by rectangles of the form $A \times B$, $A \in \SF{W}$, $B \in \SF{U}$~\cite[Pg. 97]{Ash-1972-B}.} $(\set{W}\times\set{U},\SF{W}\times\SF{U})$ via
\begin{equation*}
\rho_{\alpha\beta}\big(A \times B \big) = \rho_{\alpha}\big(A \cap G^{-1} B\big),\ A \in \SF{W},\ B \in \SF{U}\ .
\end{equation*}
The two shifts $T_{\alpha}$ and $T_{\beta}$ together define a product shift $T_{\alpha\beta}:\set{W} \times \set{U} \rightarrow \set{W} \times \set{U}$ via
$T_{\alpha\beta}\big(\mbf{w},\mbf{u}\big) = \big( T_{\alpha}\big(\mbf{w}\big),T_{\beta}\big(\mbf{u}\big)\big)$.
The combination of $\rho_{\alpha\beta}$ and $T_{\alpha\beta}$ yields a dynamical system $(\set{W} \times \set{U}$, $\SF{W} \times \SF{U}$, $\rho_{\alpha\beta}$, $T_{\alpha\beta})$.

The sequence coder $G$ is said to be \textit{$(T_{\alpha},T_{\beta})$-stationary} / \textit{$(T_{\alpha},T_{\beta})$-ergodic} / \textit{$(T_{\alpha},T_{\beta})$-AMS} if, for any $T_{\alpha}$-stationary / $T_{\alpha}$-ergodic / $T_{\alpha}$-AMS probability measure $\rho_{\alpha}$, the induced measure $\rho_{\alpha\beta}$ is $T_{\alpha\beta}$-stationary / $T_{\alpha\beta}$-ergodic / $T_{\alpha\beta}$-AMS.

\medskip

\begin{lemma}[Ex. 9.4.3, \cite{Gray-1990-B}]\label{Sec:6:Lem:Stationary-Ergodic-Coders} {\em A sequence coder $G$ is $(T_{\alpha},T_{\beta})$-stationary if and only if
$G\big(T_{\alpha}(\mbf{w})\big) = T_{\beta}\big(G(\mbf{w})\big)$.
}
\end{lemma}

\medskip

\begin{lemma}[Lems. 9.3.2 $\&$ 9.4.1, \cite{Gray-1990-B}]\label{Sec:6:Lem:AMS-Coders}
{\em If $G$ is $(T_{\alpha},T_{\beta})$-stationary, then $G$ is also
$(T_{\alpha},T_{\beta})$-ergodic and $(T_{\alpha},T_{\beta})$-AMS.}
\end{lemma}

\medskip

We note in passing that the sequence coder $F$ generated by the word function $f$ is not $(\shift{X},\shift{Y})$-stationary. Thus, Theorem~\ref{Sec:4:The:AMS-Word-Valued-Source} does not follow directly from Lemma~\ref{Sec:6:Lem:AMS-Coders}. The additional result needed to prove Theorem~\ref{Sec:4:The:AMS-Word-Valued-Source} is given in the next section.

%% file: section6d.tex
\subsection{AMS Processes $\&$ Variable Length Shifts}

Suppose that $\mbf{W}$ is a discrete-finite alphabet random process and \dynsys{W}{\rho}{W} is the corresponding dynamical system, where $\shift{W}(\mbf{w}) = w_2,w_3,\ldots$ is the left-shift transform. Now, suppose that $N$ is a natural number and $\mbf{W}$ is parsed into a sequence of non-overlapping blocks of length $N$ to form the block-valued process $\mbf{W}^N = \big\{(W_{nN+1}, W_{nN+2}, \ldots, W_{(n+1)N});\ n =  0 , 1, \ldots\big\}$. I.e. $\mbf{W}^N$ is simply $\mbf{W}$ viewed in blocks of length $N$. The appropriate shift transform for $\mbf{W}^N$ is the $N$-block shift $\shiftns{W}{}{N}: \set{W} \rightarrow \set{W}$ of Gray and Kieffer~\cite{Gray-Oct-1980-A} (see also Gray and Saadat~\cite{Gray-Jan-1984-A}), which is defined by
\begin{align*}
\shiftns{W}{}{N}(\mbf{w}) &= \shiftn{W}{N}(\mbf{w}) = w_{N+1},w_{N+2},\ldots\ .
\end{align*}
The following proposition shows that the AMS property transcends block-time scales.

\medskip

\begin{proposition}[Cor. 2.1, \cite{Gray-Jan-1984-A}]\label{Sec:6:Pro:Gray}
{\em If $\rho$ is $\shiftns{W}{}{N}$-AMS for any natural number $N$, then $\rho$ is $\shiftns{W}{}{M}$-AMS for every natural number $M$.}
\end{proposition}

\medskip

Proposition~\ref{Sec:6:Pro:Gray} does not have analogues for stationary and / or ergodic random processes; it is a unique property of AMS random processes. We now extend this proposition to include the more general notion of ``variable-length'' parsing, which will be necessary for our study of word-valued sources.

Suppose now that $\mbf{W}$ is parsed into a sequence of non-overlapping blocks, where the length of each block is determined by a simple function $\gamma: \set{W} \rightarrow \{1,2\ldots,N\}$. The appropriate transform for this variable-length parsing is the variable-length shift of Gray and Kieffer~\cite[Ex. 6]{Gray-Oct-1980-A}.

\medskip

\begin{definition}[Variable-Length Shift]\label{Sec:6:Def:Variable-Length}
Suppose that $\gamma:$ $\set{W}$ $\rightarrow$ $\{1$, $2$, $\ldots$, $N\}$ is a simple measurable function and that there exists a natural number $M$ such that $\gamma(\mbf{w}) = \gamma(\mbf{w}')$ for every pair of sequences $\mbf{w}$, $\mbf{w}'$ $\in \set{W}$ with $w_i = w'_i$ for every $i = 1,2,\ldots,M$. The variable-length shift $\shiftns{W}{}{\gamma}: \set{W} \rightarrow \set{W}$ generated by $\gamma$ is defined by~\cite{Gray-Oct-1980-A}
\begin{equation*}
\shiftns{W}{}{\gamma}(\mbf{w}) = \shiftn{W}{\gamma(\mbf{w})}(\mbf{w}) = w_{\gamma(\mbf{w})+1},w_{\gamma(\mbf{w})+2},\ldots\ .
\end{equation*}
\end{definition}

\medskip

Our extension of Proposition~\ref{Sec:6:Pro:Gray} is given in the next lemma. This lemma will be the centrepiece of our proof of Theorem~\ref{Sec:4:The:AMS-Word-Valued-Source}.

\medskip

\begin{lemma}\label{Sec:3:Lem:Variable-Length-Shift-AMS}
{\em If $\rho$ is $\shiftns{W}{}{\gamma}$-AMS for any variable-length shift $\shiftns{W}{}{\gamma}:\set{W} \rightarrow \set{W}$, then $\rho$ is $\shiftns{W}{}{\lambda}$-AMS for every variable-length shift $\shiftns{W}{}{\lambda}: \set{W} \rightarrow \set{W}$.
}
\end{lemma}

\medskip

We note that Gray's proof of Proposition~\ref{Sec:6:Pro:Gray}~\cite[Sec. 7.3]{Gray-1987-B} elegantly combines convergent subsequences with the notion of asymptotic dominance. It is not clear if this argument can be extended to prove the more general Lemma~\ref{Sec:3:Lem:Variable-Length-Shift-AMS}. Instead, we take a more laborious approach and prove the lemma by showing an ergodic theorem and applying Lemma~\ref{Sec:6:Lem:AMS-Equivalent} (iii).

\medskip

\begin{proof}
We first show that if $\rho$ is $\shiftns{W}{}{\gamma}$-AMS, then $\rho$ must also be $\shift{W}$-AMS. We then show that if $\rho$ is $\shift{W}$-AMS, then $\rho$ must also be $\shiftns{W}{}{\lambda}$-AMS.

Assume that $\rho$ is $\shiftns{W}{}{\gamma}$-AMS. From Lemma~\ref{Sec:6:Lem:AMS-Equivalent} (iv), there exists a $\shiftns{W}{}{\gamma}$-stationary probability measure $\overline{\rho}^\gamma$ on $(\set{W},\SF{W})$ such that $\shiftns{W}{-1}{\gamma} A = A$ and $\overline{\rho}^\gamma(A) = 0$ together imply that $\rho(A) = 0$. Using the procedure given by Gray and Kieffer in~\cite[Ex. 6]{Gray-Oct-1980-A}, it can be shown that $\overline{\rho}^{\gamma}$ is also $\shift{W}$-AMS. A second application of  Lemma~\ref{Sec:6:Lem:AMS-Equivalent} (iv) shows that there exists a $\shift{W}$-stationary probability measure $\overline{\rho}$ on $(\set{W},\SF{W})$ such that $\shiftn{W}{-1} A = A$ and $\overline{\rho}(A) = 0$ together imply that $\overline{\rho}^{\gamma}(A) = 0$. Note also that if a set $A$ is $\shift{W}$-invariant, then it is also $\shiftns{W}{}{\gamma}$-invariant: $A = \shiftn{W}{-1}A \Rightarrow A = \shiftns{W}{-1}{\gamma}A$. On combining these facts, we have the following: if $A = \shiftn{W}{-1}A$ and $\overline{\rho}(A) = 0$, then it must be true that $\overline{\rho}^{\gamma}(A) = 0$, $A = \shiftns{W}{-1}{\gamma}A$ and $\rho(A) = 0$. Thus, we have demonstrated the existence of a $\shift{W}$-stationary probability measure $\overline{\rho}$ on $(\set{W},\SF{W})$ such that $\shiftn{W}{-1}A = A$ and $\overline{\rho}(A) = 0$ together imply that $\rho(A) = 0$. A third application of Lemma~\ref{Sec:6:Lem:AMS-Equivalent} (iv) shows that $\rho$ must indeed be $\shift{W}$-AMS.

We now show: if $\rho$ is $\shift{W}$-AMS, then $\rho$ must also be $\shiftns{W}{}{\lambda}$-AMS. To do this, it will be useful to identify the orbit\footnote{The orbit of $\shiftns{W}{}{\lambda}$ on $\mbf{w}$ is the sequence of points $\mbf{w}$, $\shiftns{W}{}{\lambda}(\mbf{w})$, $\shiftns{W}{2}{\lambda}(\mbf{w})$, $\ldots$ from $\set{W}$.} of $\shiftns{W}{}{\lambda}$ on each sequence $\mbf{w} \in \set{W}$  with a time subsequence $\zeta = \zeta_0,\zeta_1,\ldots$. Namely, for each $n = 0,1,\ldots$ set $\zeta_n$ to be
\begin{equation}\label{Sec:6:Eqn:Time-Subsequence}
\zeta_n = \left\{
        \begin{array}{ll}
          0, & \hbox{if } n = 0\\
          \sum_{i=0}^{n-1} \lambda\Big(\shiftns{W}{i}{\lambda}(\mbf{w})\Big), & \hbox{if } n \geq 1\ ,
        \end{array}
      \right.
\end{equation}
so, by construction, we have that
\begin{equation}\label{Sec:6:Eqn:Time-Subsequence-2}
\shiftns{W}{n}{\lambda}(\mbf{w}) = w_{\zeta_n+1},w_{\zeta_n+2},\ldots = \shiftn{W}{\zeta_n}(\mbf{w})\ .
\end{equation}
Let $\xi = \xi_0,\xi_1,\ldots$ be the weight sequence that corresponds to $\zeta$, as given by~\eqref{Sec:6:Eqn:Weight-1a}. Since the length of each shift is at most $N$, the density $d_{\zeta}$ of $\zeta$ in $\mathbb{Z}^*$, as given by~\eqref{Sec:6:Eqn:Density}, can be no smaller than $1/N$ (when the limit exists).

Let $\set{U}$ denote the collection of all sequences with elements from $\{1,2,\ldots,N\}$, let $\SF{U}$ be the $\sigma$-field on $\set{U}$ generated by cylinder sets, and let $\shift{U}(\mbf{u}) = u_2,u_3,\ldots$ be the left-shift transform. Let $\Lambda: \set{W} \rightarrow \set{U}$ be the mapping defined by
\begin{equation*}
\Lambda(\mbf{w}) = \lambda\big(\mbf{w}\big),\ \lambda\big(\shift{W}(\mbf{w})\big),\  \lambda\big(\shiftn{W}{2}(\mbf{w})\big),\ \ldots\ .
\end{equation*}
From Lemma~\ref{Sec:6:Lem:Stationary-Ergodic-Coders}, this mapping is $(\shift{W},\shift{U})$-stationary since $\shift{U}(\Lambda(\mbf{w})) = \Lambda(\shift{W}(\mbf{w}))$. Finally, from Lemma~\ref{Sec:6:Lem:AMS-Coders} the induced measure $\rho_{wu}\big(A \times B\big) = \rho\big(A \cap \Lambda^{-1}B\big)$
on $(\set{W} \times \set{U},$ $\SF{W} \times \SF{U})$ is $T_{\set{W}\set{U}}$-AMS, where
$T_{\set{W}\set{U}}(\mbf{w},\mbf{u}) = \big(\shift{W}(\mbf{w}),\shift{U}(\mbf{u})\big)$.

Let $\set{Z}$ denote the collection of all sequences with elements from $\{0,1\}$, let $\SF{Z}$ be the $\sigma$-field generated by cylinder sets, and let $\shift{Z}(\mbf{z}) = z_2,z_3,\ldots$ be the left-shift transform. We now construct a finite-state coder $G: \set{W} \times \set{U} \rightarrow \set{Z}$, which identifies the orbit of the variable-length shift $\shiftns{W}{}{\lambda}$. Define $\set{G} = \{0,1,\ldots,N-1\}$ to be the internal state space of the coder, and define the state update function $g_s$ and the output function $g_o$ by
\begin{align*}
g_s(w,u,s) &= \left\{
               \begin{array}{ll}
                 u-1, & \hbox{if } s = 0 \\
                 s-1, & \hbox{otherwise.}
               \end{array}
             \right.\\
g_o(w,u,s) &= \left\{
                \begin{array}{ll}
                  1, & \hbox{if } s=0 \\
                  0, & \hbox{otherwise.}
                \end{array}
              \right.
\end{align*}
Set $s_1 = 0$ and calculate the first output $z_1 = g_o(w_1,u_1,0) = 1$. Update the state $s_2 = g_s(w_1,u_1,0) = u_1-1$ and determine the next output $z_2 = g_o(w_2,u_2,u_1-1)$. Continue in this fashion to obtain the finite state coder $G: \set{W} \times \set{U} \rightarrow \set{Z}$. As with sequence coders, the finite-state coder $G$ is measurable and it induces a probability measure
\begin{equation*}
\rho_{wuz}(A \times B \times C) = \rho_{wu}\big((A \times B)\cap G^{-1}C\big)
\end{equation*}
on $(\set{W} \times \set{U} \times \set{Z},$ $\SF{W} \times \SF{U} \times \SF{Z})$. Moreover, this finite state coder is an example of a one-sided Markov channel~\cite{Kieffer-May-1981-A}, so it follows from\footnote{Example (b) from~\cite{Kieffer-May-1981-A} demonstrates that a finite-state coder is a special case of a one-sided Markov channel.}~\cite[Thm. 6]{Kieffer-May-1981-A} that $\rho_{wuz}$ is $T_{\set{W}\set{U}\set{Z}}$-AMS, where $T_{\set{W}\set{U}\set{Z}}(\mbf{w},\mbf{u},\mbf{z}) = \big(\shift{W}(\mbf{w}),\shift{U}(\mbf{u}),\shift{Z}(\mbf{z})\big)$.

Consider the set
\begin{equation*}
\Upsilon = \big\{ (\mbf{w},\mbf{u},\mbf{z}) : \mbf{w} \in \set{W},\ \mbf{u} = \Lambda(\mbf{w}),\ \mbf{z} = G\big(\mbf{w},\Lambda(\mbf{w})\big)\big\}
\end{equation*}
It can be shown that $\Upsilon$ is measurable and $\rho_{wuz}(\Upsilon) = 1$. Suppose $(\mbf{w},\mbf{u},\mbf{z}) \in \Upsilon$, $\zeta$ is the time subsequence from~\eqref{Sec:6:Eqn:Time-Subsequence}, and $\xi$ is the weight sequence corresponding to $\zeta$. If $\mathbf{1}_{\lambda}: \set{W} \times \set{U} \times \set{Z} \rightarrow \{0,1\}$ is the indicator function defined by
\begin{equation*}
\mathbf{1}_{\lambda}(\mbf{w},\mbf{u},\mbf{z}) =
\left\{
  \begin{array}{ll}
    1, & \hbox{if } z_1 = 1 \\
    0, & \hbox{otherwise,}
  \end{array}
\right.
\end{equation*}
then, by construction, we have that
\begin{equation}
\label{Sec:6:Eqn:Weight-1}
\xi_i = \mathbf{1}_{\lambda}\big(T^i_{\set{W}\set{U}\set{Z}}(\mbf{w},\mbf{u},\mbf{z})\big)\,
\end{equation}
for all $i = 0,1,2\ldots$. Moreover, the density of $\zeta$ is given by (if the limit exists)
\begin{align}
\notag
d_{\zeta} &= \lim_{n\rightarrow\infty}\frac{1}{n}\sum_{i=0}^{n-1}\xi_i\\
\notag
&=\lim_{n\rightarrow\infty}\frac{1}{n}\sum_{i=0}^{n-1}\mathbf{1}_{\lambda}\big(T^i_{\set{W} \set{U} \set{Z}}(\mbf{w},\mbf{u},\mbf{z})\big)\\
\label{Sec:6:Eqn:Density-1}
&= \langle \mathbf{1}_{\lambda} \rangle(\mbf{w},\mbf{u},\mbf{z})\ .
\end{align}
Finally, since the length of each codeword is no more than $L$, it must be true that $d_{\zeta} \geq 1/L$ (when this limit exists.)

Since $\rho_{wuz}$ is $T_{\set{W} \set{U} \set{Z}}$-AMS, it follows from Lemma~\ref{Sec:6:Lem:AMS-Equivalent} (iii) that there exists a subset $\Omega$ with probability $\rho_{wuz}(\Omega) = 1$ such that, for each $(\mbf{w},\mbf{u},\mbf{z}) \in \Omega$, the limit
\begin{equation*}
\langle g \rangle(\mbf{w},\mbf{u},\mbf{z}) = \lim_{n\rightarrow\infty}\frac{1}{n}\sum_{i=0}^{n-1}
g\big(T_{\set{W} \set{U} \set{Z}}^i(\mbf{w},\mbf{u},\mbf{z})\big)
\end{equation*}
exists for every bounded-measurable $g$. Since $\mathbf{1}_{\lambda}$ is bounded and measurable, this ergodic theorem guarantees the density~\eqref{Sec:6:Eqn:Density-1} exists for every $(\mbf{w},\mbf{u},\mbf{z}) \in \Omega \cap \Upsilon$.

Let $T_{\set{W}\set{U}\set{Z}^{\lambda}}$ denote the variable-length shift on the product space $\set{W} \times \set{U} \times \set{V}$ defined by
\begin{equation*}
T_{\set{W}\set{U}\set{Z}^{\lambda}}(\mbf{w},\mbf{u},\mbf{z}) = T_{\set{W}\set{U}\set{Z}}^{\lambda(\mbf{w})}(\mbf{w},\mbf{u},\mbf{z})\ .
\end{equation*}
From~\eqref{Sec:6:Eqn:Time-Subsequence}, we have that $T^n_{\set{W}\set{U}\set{Z}^{\lambda}}(\mbf{w},\mbf{u},\mbf{z}) = T_{\set{W}\set{U}\set{Z}}^{\zeta_n}(\mbf{w},\mbf{u},\mbf{z})$ for all $n = 0,1,2\ldots$.

If $g: \set{W} \times \set{U} \times \set{Z} \rightarrow (-\infty,\infty)$ is bounded-measurable, then $\mathbf{1}_{\lambda} \times g$ is bounded and measurable, and for each $(\mbf{w},\mbf{u},\mbf{z}) \in \Omega \cap \Upsilon$ the following limits will exist:
\begin{align}
\notag
\langle \mathbf{1}_{\lambda} \times g \rangle
&=\lim_{n\rightarrow\infty}\frac{1}{n}\sum_{i=0}^{n-1}
\mathbf{1}_{\lambda}\big(T^i_{\set{W}\set{U}\set{Z}}(\mbf{w},\mbf{u},\mbf{z})\big)
g\big(T^i_{\set{W}\set{U}\set{Z}}(\mbf{w},\mbf{u},\mbf{z})\big)\\
\label{Sec:6:Eqn:Limit-1}
&=\lim_{n\rightarrow\infty}\frac{1}{n}\sum_{i=0}^{n-1}
\xi_i
g\big(T^i_{\set{W}\set{U}\set{Z}}(\mbf{w},\mbf{u},\mbf{z})\big)\\
\label{Sec:6:Eqn:Limit-2}
&=d_{\zeta}\ \ \lim_{m\rightarrow\infty}\frac{1}{m}\sum_{j=0}^{m-1}
g\big(T^{\zeta_j}_{\set{W}\set{U}\set{Z}}(\mbf{w},\mbf{u},\mbf{z})\big)\\
\label{Sec:6:Eqn:Limit-3}
&=d_{\zeta}\ \ \lim_{m\rightarrow\infty}\frac{1}{m}\sum_{j=0}^{m-1}
g\big(T^j_{\set{W}\set{U}\set{Z}^{\lambda}}(\mbf{w},\mbf{u},\mbf{z})\big)\ ,
\end{align}
where~\eqref{Sec:6:Eqn:Limit-1} follows from~\eqref{Sec:6:Eqn:Weight-1}, \eqref{Sec:6:Eqn:Limit-2} follows from Lemma~\ref{Sec:6:Lem:Bellow}, and~\eqref{Sec:6:Eqn:Limit-3}  follows from~\eqref{Sec:6:Eqn:Time-Subsequence}.
This chain of equalities guarantees the limit in \eqref{Sec:6:Eqn:Limit-3} exists for every $(\mbf{w},\mbf{u},\mbf{z}) \in \Omega \cap \Upsilon$. Since $g$ is an arbitrary bounded measurable function, it follows from Lemma~\ref{Sec:6:Lem:AMS-Equivalent} (iii) that $\rho_{wuz}$ is $T_{\set{W}\set{U}\set{Z}^\lambda}$-AMS. Finally, since $\rho$ is a marginal of $\rho_{wuz}$, it follows that $\rho$ is $\shiftns{W}{}{\lambda}$-AMS.
\end{proof} 

%% file: section6e.tex
\subsection{Ergodic Processes $\&$ Variable Length Shifts}\label{Sec:6:Sub:E}

In Lemma~\ref{Sec:3:Lem:Variable-Length-Shift-AMS}, it was shown that an AMS random process remains AMS under all variable-length time shifts. The next lemma proves a weaker result for ergodic processes. Again, suppose that $\mbf{W}$ is a discrete-finite alphabet random process and \dynsys{W}{\rho}{W} is the corresponding dynamical system.

\medskip

\begin{lemma}\label{Sec:3:Lem:Variable-Length-Shift-Erogidc}
{\em If $\rho$ is $\shiftns{W}{}{\gamma}$-ergodic for some variable-length shift $\shiftns{W}{}{\gamma}: \set{W} \rightarrow \set{W}$, then $\rho$ is also $\shift{W}$-ergodic.}
\end{lemma}

\medskip

\begin{proof}
If $\rho$ is $\shiftns{W}{}{\gamma}$-ergodic and $A$ is an $\shiftns{W}{}{\gamma}$-invariant set, then $\rho(A) = 0$ or $1$. Since $A = \shiftn{W}{-1}A$ implies that $A = \shiftns{W}{-1}{\gamma} A$, it follows that $\rho(A) = 0$ or $1$ for every $\shift{W}$-invariant set $A$. 
\end{proof} 

%% file: section6f.tex
\subsection{Proof of Theorem~\ref{Sec:4:The:AMS-Word-Valued-Source} (Forward Claim)}\label{Sec:6:Sub:F}

We now prove the forward claim of Theorem~\ref{Sec:4:The:AMS-Word-Valued-Source}: if $\mu$ is $\shift{X}$-AMS (and $\shift{X}$-ergodic), then $\eta$ is $\shift{Y}$-AMS (and $\shift{Y}$-ergodic). Let $\set{Z}$ denote the set of all sequences with elements from $\{1,2,\ldots,N\}$, let $\SF{Z}$ denote the $\sigma$-field generated by cylinder sets, and let $\shift{Z}(\mbf{z}) = z_2,z_3,\ldots$ denote the left-shift transform. Using the word function $f$, define the mapping
\begin{equation*}
\tilde{f}(x) = \big(f(x)_1,|f(x)|\big),\big(f(x)_2,|f(x)|-1\big),\ldots,\big(f(x)_{|f(x)|},1\big)\ ,
\end{equation*}
where $f(x)_j$, $1\leq j \leq |f(x)|$, denotes the $j^{th}$ symbol of the codeword $f(x)$. By construction, $\tilde{f}(x)$ couples the codeword $f(x)$ with a sequence of indices $|f(x)_1|,|f(x)_1|-1,\ldots,1$, which mark the distance from the current symbol to the end of the codeword. Using $\tilde{f}$, define the sequence coder $\tilde{F}: \set{X} \rightarrow \set{Y} \times \set{Z}$ via $\tilde{F}(\mbf{x}) = \tilde{f}(x_1),\tilde{f}(x_2),\ldots$.
As before, this sequence coder induces a probability measure $\eta_{yz}(A \times B) = \mu\big(\tilde{F}^{-1}(A\times B)\big)$
on $(\set{Y}\times\set{Z},$ $\SF{Y}\times\SF{Z})$. Let $T_{\set{Y}\set{Z}}(\mbf{y},\mbf{z}) = \big(\shift{Y}(\mbf{y}),\shift{Z}(\mbf{z})\big)$,
and let $T_{\set{Y}\set{Z}^\gamma}$ be the variable-length shift defined by setting $\gamma(\mbf{y},\mbf{z}) = z_1$.
Since
\begin{equation*}
\tilde{F}\big(\shift{X}(\mbf{x})\big) = T_{\set{Y}\set{Z}^\gamma}\big(\tilde{F}(\mbf{x})\big)\ .
\end{equation*}
it follows from Lemma~\ref{Sec:6:Lem:Stationary-Ergodic-Coders} that $\tilde{F}$ is a $(\shift{X},T_{\set{Y}\set{Z}^\gamma})$-stationary sequence coder. Since $\mu$ is $\shift{X}$-AMS (and $\shift{X}$-ergodic), we have from Lemma~\ref{Sec:6:Lem:AMS-Coders} that $\eta_{yz}$ is $T_{\set{Y}\set{Z}^\gamma}$-AMS (and $T_{\set{Y}\set{Z}^\gamma}$-ergodic). Finally, from Lemmas~\ref{Sec:3:Lem:Variable-Length-Shift-AMS} and~\ref{Sec:3:Lem:Variable-Length-Shift-Erogidc}, we can see that $\eta_{yz}$ must also be $T_{\set{Y}\set{Z}}$-AMS (and $T_{\set{Y}\set{Z}}$-ergodic); therefore, $\eta$ must be $\shift{Y}$-AMS (and $\shift{Y}$-ergodic).

%% file: section6g.tex
\subsection{Proof of Theorem~\ref{Sec:4:The:AMS-Word-Valued-Source} (Reverse Claim)}\label{Sec:6:Sub:G}

We now prove the reverse claim of Theorem~\ref{Sec:4:The:AMS-Word-Valued-Source}: if $\eta$ is $\shift{Y}$-AMS and $f$ is prefix-free, then $\mu$ is $\shift{X}$-AMS. Define the variable-length shift $\shiftns{Y}{}{\gamma}: \set{Y} \rightarrow \set{Y}$ by setting
\begin{equation*}
\gamma(\mbf{y}) = \left\{
                   \begin{array}{ll}
                     |c|, & \hbox{if there exists a unique $c \in \set{C}$ such that $y_i = c_i$} \\
                            &\hbox{for all $i = 1,2,\ldots,|c|$.}\\
                     1, & \hbox{otherwise.}
                   \end{array}
                 \right.
\end{equation*}
From Lemma~\ref{Sec:6:Lem:AMS-Coders}, it follows that $\eta$ is $\shiftns{Y}{}{\gamma}$-AMS.

Define
\begin{equation*}
\Omega = \big\{ \mbf{y} \in \set{Y} : \text{ there exists } \mbf{x} \in \set{X} \text{ such that } \mbf{y} = F(\mbf{x}) \big\}\ ,
\end{equation*}
where it can be shown that $\Omega \in \SF{Y}$ and $\eta(\Omega) = 1$.

Let $g: \set{C} \rightarrow \set{A}$ denote the inverse of $f$. If $\mbf{y}$ is in $\Omega$, then there exists a unique sequence of codewords $c_1,c_2,\ldots$ from $\set{C}$ such that $\mbf{y} = c_1,c_2,\ldots$. Therefore, using $g$, we may define the sequence-coder $G: \Omega \rightarrow \set{X}$ by setting $G(\mbf{y}) = F^{-1}(c_1,c_2,\ldots) = g(c_1),g(c_2),\ldots$.

For each $\mbf{y} \in \Omega$ we have that
$
G\big(\shiftns{Y}{}{\gamma}(\mbf{y})\big) = \shift{X}\big(G(\mbf{y})\big)\ ,
$
so it follows from Lemma~\ref{Sec:6:Lem:Stationary-Ergodic-Coders} that $G$ is a $(\shiftns{Y}{}{\gamma},\shift{X})$-stationary sequence coder. From Lemma~\ref{Sec:6:Lem:Stationary-Ergodic-Coders}, the induced probability measure $\tilde{\mu}(A) = \eta(G^{-1} A)$ on $(\set{X},\SF{X})$  is $\shift{X}$-AMS. Since $\tilde{\mu}(A) = \eta(G^{-1}A) = \mu(F^{-1}G^{-1}A) = \mu(A)$ for each $A \in \SF{X}$, it follows that $\mu$ is $\shift{X}$-AMS. \hfill $\QED$

%% file: section7.tex
\section{Proof of Theorem~\ref{Sec:5:The:WVS-AEP} $\&$ Corollaries}\label{Sec:7}

\input{./section7a}
\input{./section7b}

%% file: section7a.tex
\subsection{Proof of Theorem~\ref{Sec:5:The:WVS-AEP}}

Let $\{\overline{\mu}_{\mbf{x}} : \mbf{x} \in \set{X}\}$ and $\{\overline{\eta}_{\mbf{y}} : \mbf{y} \in \set{Y}\}$ be the ergodic decompositions of $(\set{X},\SF{X})$ and $(\set{Y},\SF{Y})$ respectively. For each $n = 1,2,\ldots$, let $\phi_n: \set{Y} \rightarrow \set{B}^n$ be the projection $\phi_n(\mbf{y}) = y_1,y_2,\ldots,y_n$. From Lemma~\ref{Sec:5:Lem:AMS-AEP}, there exists a subset $\Omega_{x,1} \in \SF{X}$ with probability $\mu(\Omega_{x,1}) = 1$ such that the sample-entropy rate of each sequence $\mbf{x} \in \Omega_{x,1}$ exists and is given by $h(\mu,\mbf{x}) = \varphi_x(\mbf{x})$, where  $\varphi_x(\mbf{x}) = \overline{H}(\overline{\mu}_{\mbf{x}})$. Similarly, there exists a subset $\Omega_y \in \SF{Y}$ with probability $\eta(\Omega_y) = 1$ such that the sample-entropy rate of each sequence $\mbf{y} \in \Omega_y$ exists and is given by $h(\eta,\mbf{y}) = \varphi_y(\mbf{y})$, where $\varphi_y(\mbf{y}) = \overline{H}(\overline{\eta}_{\mbf{y}})$. Finally, from Lemma~\ref{Sec:5:Lem:Ergodic-Decomposition} there exists a subset $\Omega_{x,2} \in \SF{X}$ with probability $\mu(\Omega_{x,2}) = 1$ such that for each sequence $\mbf{x} \in \Omega_{x,2}$ the time-averaged codeword-length exists and is given by
\begin{align*}
\lim_{n\rightarrow\infty}\frac{1}{n} \sum_{i=1}^{n}|f(x_{i})|
&= \lim_{n\rightarrow\infty}\frac{1}{n} \sum_{i=0}^{n-1}l(\shiftn{X}{i}(\mbf{x}))= \mathbb{E}\big[\overline{\mu}_{\mbf{x}},l\big]\ .
\end{align*}

For each $\mbf{x} \in \set{X}$, define the time subsequence $\zeta = \zeta_0,\zeta_1,\ldots$ by setting
\begin{equation*}
\zeta_n = \left\{
        \begin{array}{ll}
          0, & \hbox{if } n = 0 \\
          \sum_{i=1}^n|f(x_i)|, & \hbox{if } n \geq 1\ .
        \end{array}
      \right.
\end{equation*}
For each $n = 1,2,\ldots$, we have that $F^{-1}[\phi_{\zeta_n}(F(\mbf{x}))] \supseteq [x^n]$,
with set equality if $f$ is prefix free. This implies
\begin{equation}\label{Sec:7:Eqn:AEP-1}
\frac{1}{n} \log_2 \frac{1}{\mu\big([x^n]\big)} \geq \frac{\zeta_n}{n} \frac{1}{\zeta_n} \log_2 \frac{1}{\eta\Big(\Big[\phi_{\zeta_n}\big(F(\mbf{x})\big)\Big]\Big)}\ ,
\end{equation}
with equality if $f$ is prefix free. Furthermore,
\begin{equation}\label{Sec:7:Eqn:Sample-Entropy-Subsequence}
\frac{1}{\zeta_n} \log_2 \frac{1}{\eta\Big(\big[\phi_{\zeta_n}\big(F(\mbf{x})\big)\big]\Big)},\  n = 1,2,\ldots,
\end{equation}
is a subsequence of
\begin{equation}\label{Sec:7:Eqn:Sample-Entropy}
\frac{1}{n} \log_2 \frac{1}{\eta\Big(\big[\phi_n\big(F(\mbf{x})\big)\big]\Big)},\ n = 1,2,\ldots\ ;
\end{equation}
thus, if $\mbf{x} \in F^{-1} \Omega_y$, then \eqref{Sec:7:Eqn:Sample-Entropy-Subsequence} and \eqref{Sec:7:Eqn:Sample-Entropy} both converge to $\varphi_{y}(F(\mbf{x}))$ as $n \rightarrow \infty$. To complete the proof, note that Theorem~\ref{Sec:5:The:WVS-AEP} follows from~\eqref{Sec:7:Eqn:AEP-1} since $\lim_{n\rightarrow\infty} \zeta_n / n = \mathbb{E}[\overline{\mu}_{\mbf{x}},l]$, $\lim_{n\rightarrow\infty}-(1/n) \log_2 \mu([x^n]) = \overline{H}(\overline{\mu}_{\mbf{x}})$ and $\lim_{n\rightarrow\infty} -(1/n) \log_2 \eta([\phi_{\zeta_n}(F(\mbf{x}))])$ exists for every $\mbf{x} \in \Omega_{x,1} \cap \Omega_{x,2} \cap F^{-1} \Omega_y$. \hfill $\QED$ 

%% file: section7b.tex
\subsection{Proof of Corollary~\ref{Sec:5:The:WVS-AEP:Cor:1}}

Let $\{\overline{\mu}_{\mbf{x}} : \mbf{x} \in \set{X}\}$ and $\{\overline{\eta}_{\mbf{y}} : \mbf{y} \in \set{Y}\}$ be the ergodic decompositions of $(\set{X},\SF{X})$ and $(\set{Y},\SF{Y})$ respectively. As usual, define $\varphi_x(\mbf{x}) = \overline{H}(\overline{\mu}_{\mbf{x}})$ and $\varphi_y(\mbf{y}) = \overline{H}(\overline{\eta}_{\mbf{y}})$. Now define $\tilde{\varphi}_x(\mbf{x}) = \varphi_y(F(\mbf{x}))$ and
\begin{equation*}
g(\mbf{x}) = \frac{\varphi_{\mbf{x}}(\mbf{x})}{\mathbb{E}[\overline{\mu}_{\mbf{x}},l]}\ .
\end{equation*}

Suppose $\mu$ is $\shift{X}$-AMS. From Theorem~\ref{Sec:5:The:WVS-AEP}, we have that $\eta$ is $\shift{Y}$-AMS and $\tilde{\varphi}_x(\mbf{x}) \leq g(\mbf{x})$ on a set $\Omega_x$ of probability $\mu(\Omega_x) = 1$ (with equality if $f$ is prefix-free). Therefore,
\begin{equation}\label{Sec:7:Eqn:Int-1}
\int \tilde{\varphi}_x(\mbf{x})\ d \mu(\mbf{x}) \leq \int g(\mbf{x})\ d \mu(\mbf{x})\ .
\end{equation}
Note, the R.H.S. of~\eqref{Sec:7:Eqn:Int-1} is equal to the R.H.S. of~\eqref{Sec:2:Eqn:WVS-AEP-2}. By the change of variables
formula~\cite[Lem. 4.4.7]{Gray-1987-B} and Lemma~\ref{Sec:5:Lem:AMS-AEP}, we have
\begin{equation}\label{Sec:7:Eqn:Int-2}
\int \tilde{\varphi}_x(\mbf{x})\ d\mu(\mbf{x}) = \int \varphi_y(\mbf{y})\ d\eta(\mbf{y})=\overline{H}(\eta)\  .
\end{equation}
which is the desired result. \hfill $\QED$

\subsection{Proof of Corollary~\ref{Sec:5:The:WVS-AEP:Cor:2}}

Suppose that $\mu$ is $\shift{X}$-stationary and $\shift{X}$-ergodic. From Theorem~\ref{Sec:4:The:AMS-Word-Valued-Source}, $\eta$ is $\shift{Y}$-ergodic. From Lemma~\ref{Sec:5:Lem:AMS-AEP}, there exists a subset $\Omega_y \in \SF{Y}$ with probability $\eta(\Omega_y) = 1$ such that the sample-entropy rate of each sequence $\mbf{y} \in \Omega_y$ takes the same constant value $h(\eta,\mbf{y}) = \overline{H}(\eta)$. From Theorem~\ref{Sec:5:The:WVS-AEP}, there exists a subset $\Omega_x \in \SF{X}$ with probability $\mu(\Omega_x) = 1$ such that the sample-entropy rate of each coded sequence $F(\mbf{x})$, $\mbf{x} \in \Omega_x$, exists and is bound from above by
\begin{equation}\label{Sec:7:Eqn:Y-AEP-1-a}
h(\eta,F(\mbf{x})) \leq \frac{\overline{H}(\overline{\mu}_{\mbf{x}})}{\mathbb{E}\big[\overline{\mu}_{\mbf{x}},l\big]}\ .
\end{equation}

Since $F^{-1}\Omega_y \cap \Omega_x \neq \emptyset$, there exists $\mbf{x} \in \Omega_x$ and $\mbf{y} \in \Omega_y$ such that $\mbf{y} = F(\mbf{x})$ and
\begin{equation}\label{Sec:7:Eqn:Y-AEP-1}
h\big(\eta,\mbf{y}\big)
\leq
\frac{\overline{H}(\overline{\mu}_{\mbf{x}})}
{\mathbb{E}\big[\overline{\mu}_{\mbf{x}},l\big]} = \frac{\overline{H}(\mu)}
{\mathbb{E}\big[\mu,l\big]}\
\end{equation}
where the R.H.S. equality in~\eqref{Sec:7:Eqn:Y-AEP-1} follows from the fact that $\mu$ is $\shift{X}$-stationary and $\shift{X}$-ergodic. The result follows since $h\big(\eta,\mbf{y}\big)$ exists and takes the constant value $\overline{H}(\eta)$ on $\Omega_y$. Finally, note that for prefix-free codes~\eqref{Sec:7:Eqn:Y-AEP-1-a}  and therefore~\eqref{Sec:7:Eqn:Y-AEP-1} are equalities. \hfill $\QED$

%% file: acknowledgements.tex
\section*{Acknowledgements}
%Acknowledgements will appear in the final version.

The authors are indebted to Alex Grant, Ingmar Land, Oliver Nagy and the two anonymous reviewers for their thoughtful comments on the manuscript. These comments have greatly improved its quality.